\documentclass[12pt,preprint]{aastex}

\def\sn{{SN~2005ke}}

\def\C{{\sl Chandra}}

\def\S{{\sl Swift}}

\def\gs{\mathrel{\mathchoice {\vcenter{\offinterlineskip\halign{\hfil
$\displaystyle##$\hfil\cr>\cr\sim\cr}}}
{\vcenter{\offinterlineskip\halign{\hfil$\textstyle##$\hfil\cr
>\cr\sim\cr}}}
{\vcenter{\offinterlineskip\halign{\hfil$\scriptstyle##$\hfil\cr
>\cr\sim\cr}}}
{\vcenter{\offinterlineskip\halign{\hfil$\scriptscriptstyle##$\hfil\cr
>\cr\sim\cr}}}}}
\def\ls{\mathrel{\mathchoice {\vcenter{\offinterlineskip\halign{\hfil
$\displaystyle##$\hfil\cr<\cr\sim\cr}}}
{\vcenter{\offinterlineskip\halign{\hfil$\textstyle##$\hfil\cr
<\cr\sim\cr}}}
{\vcenter{\offinterlineskip\halign{\hfil$\scriptstyle##$\hfil\cr
<\cr\sim\cr}}}
{\vcenter{\offinterlineskip\halign{\hfil$\scriptscriptstyle##$\hfil\cr
<\cr\sim\cr}}}}}

\begin{document}

\title{X-Ray Observations of Type Ia Supernovae with {\sl Swift}: \\
Evidence for Circumstellar Interaction for SN~2005\lowercase{ke}}

\author{S.~Immler\altaffilmark{1,2},
P. J. Brown\altaffilmark{3}, 
P. Milne\altaffilmark{4}, 
L.-S. The\altaffilmark{5}, 
R. Petre\altaffilmark{1}, 
N. Gehrels\altaffilmark{1}, 
D. N. Burrows\altaffilmark{3}, 
J. A. Nousek\altaffilmark{3}, 
C. L. Williams\altaffilmark{1}, 
E. Pian\altaffilmark{6},
P. A. Mazzali\altaffilmark{6,7},
K. Nomoto\altaffilmark{8},
R. A. Chevalier\altaffilmark{9},
V. Mangano\altaffilmark{10}, 
S. T. Holland\altaffilmark{1,2}, 
P. W. A. Roming\altaffilmark{3},
J. Greiner\altaffilmark{11} and 
D. Pooley\altaffilmark{12,13}
}

\altaffiltext{1}{Exploration of the Universe Division, 
X-Ray Astrophysics Laboratory, Code 662, 
NASA Goddard Space Flight Center, Greenbelt, MD 20771, USA}
\altaffiltext{2}{Universities Space Research Association, 
10211 Wincopin Circle, Columbia, MD 21044, USA}
\altaffiltext{3}{Department of Astronomy and Astrophysics, Pennsylvania 
State University, 525 Davey Laboratory, University Park, PA 16802, USA}
\altaffiltext{4}{Steward Observatory, 933 North Cherry Avenue, RM N204 Tucson, 
AZ 85721, USA}
\altaffiltext{5}{Dept. of Physics and Astronomy, Clemson University, Clemson, 
SC 29634-0978, USA}
\altaffiltext{6}{INAF, Osservatorio Astronomico di Trieste, via G.B. 
Tiepolo 11, 34131 Trieste, Italy}
\altaffiltext{7}{Max-Planck-Institut f\"ur Astrophysik, Karl-Schwarzschild
Strasse 1, 85741 Garching, Germany}
\altaffiltext{8}{Department of Astronomy, School of Science, University of 
Tokyo, Bunkyo-ku, Tokyo 113-0033, Japan}
\altaffiltext{9}{Department of Astronomy, University of Virginia, P.O. Box 3818, Charlottesville, VA 22903, USA}
\altaffiltext{10}{INAF-Istituto di Astrofisica Spaziale e Cosmica, 
Via Ugo La Malfa 153, I-90146 Palermo, Italy}
\altaffiltext{11}{Max-Planck-Institut f\"ur extraterrestrische Physik, 
Giessenbachstrasse, 85748 Garching, Germany}
\altaffiltext{12}{Astronomy Department, University of California, Berkeley, 
601 Campbell Hall, Berkeley, CA 9472, USA}
\altaffiltext{13}{Chandra Fellow}

\shorttitle{X-Ray Observations of Type Ia SNe with {\sl Swift}}
\shortauthors{Immler et~al.}

\begin{abstract}

We present a study of the early (days to weeks) X-ray and UV properties 
of eight Type Ia supernovae (SNe Ia) which have been extensively observed 
with the X-Ray Telescope (XRT) and UV/Optical Telescope (UVOT) onboard \S, 
ranging from 5--132 days after the outburst.
SN~2005ke is tentatively detected (at a 3--3.6$\sigma$ level of significance) 
in X-rays based on deep monitoring with the XRT ranging from 8 to 120 days after 
the outburst. 
The inferred X-ray luminosity [$L_{0.3-2}=(2\pm1)\times10^{38}~{\rm ergs~s}^{-1}$; 
0.3--2~keV band] is likely caused by interaction of the SN shock with circumstellar 
material (CSM), deposited by a stellar wind from the progenitor's companion star 
with a mass-loss rate of 
$\dot{M} \approx 3 \times 10^{-6}~M_{\odot}~{\rm yr}^{-1}~(v_{\rm w}/10~{\rm km~s}^{-1})$.
Evidence of CSM 
interaction in X-rays is independently confirmed by an excess of UV emission as 
observed with the UVOT onboard \S, starting around 35 days after the explosion.
The non-detection of \sn\ with \C\ 105~days after the outburst implies a rate of 
decline steeper than $L_{\rm x} \propto t^{-0.75}$, consistent with the decline
expected from the interaction of the SN shock with a spherically symmetric CSM
($t^{-1}$).
None of the other seven SNe Ia is detected in X-rays or shows a UV excess, which 
allows us to put tight constraints on the mass-loss rates of the progenitor systems.

\end{abstract}


\section{Introduction}
\label{introduction}

Type Ia supernovae (SNe Ia) are a subclass of exploding stars defined observationally 
by the absence of hydrogen lines in their optical spectra and the presence of lines 
from elements such as silicon and sulfur (Leibundgut 2000). There is consensus that 
SNe Ia are explosions of white dwarfs which occur when accretion from a companion 
star drives the white dwarf mass close to the Chandrasekhar limit (Woosley \& Weaver 1986,
Nomoto et~al.\ 2003). However, the details of the system are not fully understood,
especially with regard to the type of companion (a main sequence star, a red giant). 
A useful indicator of the properties of the companion is through its mass loss, which
depends strongly on the stellar type. The X-rays and the radio regimes are especially 
well suited to study the interaction of the SN ejecta with the surrounding CSM, which
should be dominated by the companion's wind. However, no SN Ia has ever been detected 
in either regime.

In this Letter, we present X-ray and UV data for a sample of eight SNe~Ia observed 
with \S\ between 5 and 132 days after outburst. The dates of outburst were estimated to 
be $18\pm2$ days before the peak in the $B$-band. The highlight of this study is the 
tentative detection of SN~2005ke in X-rays with \S, which was discovered on 
2005-11-13.33~UT with the Katzman Automatic Imaging Telescope (KAIT; Baek, Prased 
\& Li 2005) and later classified as an under-luminous SN Ia from the presence of  
the characteristic 420~nm Ti~II and 635~nm Si~II lines (Patat \& Baade 2005). 

\section{Observations and Data Reduction}
\label{obs}

A sample of eight SNe~Ia was observed with the XRT (Burrows et~al.\ 2005) and UVOT 
(Roming et~al.\ 2005) instruments onboard the \S\ Observatory (Gehrels et~al.\ 2004), 
ranging from days to weeks after the outburst. Multiple exposures were obtained in 
the $V$, $B$, $U$, $UVW1$ (181--321~nm), $UVM2$ (166--268~nm), and $UVW2$ (112--264~nm)
UVOT filters and with the XRT (0.2--10 keV band) in photon counting (PC) mode. 

The \S\ data were analyzed using the 
HEASOFT\footnote{http://heasarc.gsfc.nasa.gov/docs/software/lheasoft/} 
(version 6.04) and \S\ Software (version 2.3, build 17) tools and latest calibration 
products. X-ray counts were extracted from a circular region with an aperture of 
$10''$ radius centered at the optical positions of the SNe. The background was 
extracted locally from two neighboring circular regions of $10''$ radius located at 
the same distances from the nuclei of the host galaxies to account for residual 
diffuse emission from the galaxies. 

A 14.8~ks \C\ ACIS-S Director's Discretionary Time (DDT) observation of \sn\
was obtained on 2006-02-19 (sequence 500693), corresponding to day 105 after the 
outburst. The data were analyzed with CIAO\footnote{http://cxc.harvard.edu/ciao/}
(version 3.3.0.1) and the latest calibration products.

\section{Properties of the Sample}
\label{results}

This \S\ data set on SNe Ia is unique as it represents the best sample of any 
SNe Ia observed in the UV. 
While the UV lightcurves are strikingly similar (see Fig.~1), \sn\ shows a 
significant UV excess starting around day 35 after the outburst. A detailed 
analysis of the UV lightcurves will be presented in Milne et~al.\ (in preparation).

The upper limits to the X-ray luminosities of SNe 2005am, 2005cf, 2005df, 2005hk, 
2006E, and 2006X are a few $\times 10^{39}~{\rm ergs~s}^{-1}$, implying 
mass-loss rates lower than a few $\times 10^{-5}~M_{\odot}~{\rm yr}^{-1}$ for 
thermal emission (see Tab.~1). The mass-loss rate of SN~2005gj is not well 
constrained owing to the relatively short XRT exposure time (5~ks) and the large 
distance to the SN (265~Mpc). While the above upper limits are consistent with the
X-ray luminosity of \sn, the SN is peculiar as it is the only SN Ia of our sample
showing excess UV emission after peak.



\section{Supernova 2005\lowercase{ke}}
\label{discussion}

Forty-three individual exposures of \sn\ were obtained between 8 and 120 
days after the explosion (2005-11-06$\pm 2$ days). X-ray and UV images 
of SN~2005ke and its host galaxy NGC~1371 are shown in Fig.~2.

An excess of X-ray counts is detected from the position of SN~2005ke
(3--3.6$\sigma$ level of significance depending on the location of the 
background extraction regions) in the merged XRT data ranging from day 
1--54 after the outburst, with a point-spread 
function, sampling deadtime and vignetting-corrected net count rate of 
$(2.2\pm0.6) \times 10^{-4}~{\rm cts~s}^{-1}$. 
Adopting a thermal plasma spectrum with a temperature of $kT = 10$~keV 
(see Fransson, Lundqvist \& Chevalier 1996 and references therein) and assuming a 
Galactic foreground column density with no intrinsic absorption 
($N_{\rm H} = 6.16 \times 10^{20}~{\rm cm}^{-2}$; Dickey \& Lockman 1990)
we obtain a 0.3--2 keV X-ray band flux and luminosity of 
$f_{0.3-2} = (4 \pm 1) \times 10^{-15}~{\rm ergs~cm}^{-2}~{\rm s}^{-1}$ and 
$L_{0.3-2} = (2 \pm 1) \times 10^{38}~{\rm ergs~s}^{-1}$, 
respectively, for a distance of 20.7~Mpc ($z = 0.00488$, 
$H_0 = 71~{\rm km~s}^{-1}~{\rm Mpc}~^{-1}$, $\Omega_{\Lambda}=2/3$,  
$\Omega_{\rm M}=1/3$; Koribalski et~al.\ 2004). 
The chance probability of a background AGN at the measured flux being within a 
radius of $10''$ of the position of SN~2005ke is estimated to be $\ls 10^{-3}$ 
(Hasinger et~al.\ 2001).

No X-ray source is visible at the position of SN~2005ke in the \C\ observation.
The inferred $3\sigma$ upper limit to the X-ray count rate of 
$<3.5 \times 10^{-4}~{\rm cts~s}^{-1}$ corresponds to a flux and luminosity of 
$f_{0.3-2}<2.3\times10^{-15}~{\rm ergs~cm}^{-2}~{\rm s}^{-1}$ and
$L_{0.3-2}<1.2\times10^{38}~{\rm ergs~s}^{-1}$, respectively, using the same
spectral template above.

The \S\ XRT X-ray luminosity inferred for SN~2005ke is lower 
than the lowest published upper limit of the early X-ray emission of SNe Ia 
(e.g., $5 \times 10^{38}~{\rm ergs~s}^{-1}$ for SN 1992A on day 16; Schlegel \& 
Petre 1993). This improvement in the sensitivity of our measurement allows us to 
place significant constraints on SN Ia models.
The SN, however, is not detected at late times (day 105) with \C.
If the X-ray source detected with \S\ is indeed the SN, a rate of decline of
$L_{\rm x} \propto t^{n}$ with index $n<-0.75$ ($3\sigma$) is inferred.  
This is consistent with the X-ray rate of decline expected from the interaction
of the SN shock with a spherically symmetric CSM ($t^{-1}$ for thermal emission), 
as well as observations of X-ray emitting core-collapse SNe (see Immler \& Lewin 
2003 and references therein).
No other X-ray source or enhancements of the diffuse emission within the 
host galaxy is visible in the high-resolution ($\ls 1''$ FWHM ) \C\ images 
within the XRT counts extraction aperture ($10''$ radius), which excludes the
possibility that an X-ray binary or clumps in the diffuse emission in NGC~1371 
might have caused the excess emission observed with the \S\ XRT.
Each of the other $>3\sigma$ enhancements in the XRT image, however, have
matching \C\ X-rays sources.

In the following we therefore assume that the \S\ detection of SN~2005ke in 
X-rays is real, but even if it is only an upper limit, our analyses hold true. 

Our detection implies either X-ray emission from Compton-scattered $\gamma$-rays 
of the radioactive decay products of the SN ejecta or an interaction of the SN 
shock with a sufficiently dense CSM. For comparison of the observed X-ray luminosity 
(or upper limit) with the expected X-ray luminosity from SN Ia models, we calculated 
the emergent hard X-ray spectrum of model W7 (Nomoto, Thielemann, \& Yokoi 1994) 
using a Monte Carlo $\gamma$-ray transport code to simulate the propagation of 
photons produced by the radioactive-decay of $^{56}$Ni and $^{56}$Co and 
experiencing Compton scattering, photo-electric 
absorption, and pair production processes (Burrows \& The 1990).
Bremsstrahlung emission is expected to dominate the radiation in the 
0.3--2.0~keV band (Clayton \& The 1991). We calculate the bremsstrahlung 
emergent spectrum of model W7 (as in Clayton \& The 1991) but with much improved 
statistics. The bremsstrahlung 
component extends to $\approx 0.2$~keV at $t = 12$ days and lower at later times. 
The total luminosities in the energy range of 
0.3--2.0~keV are $9.2 \times 10^{33}$, $1.0 \times 10^{34}$, and 
$1.6 \times 10^{34}~{\rm  ergs~s}^{- 1}$ at $t = 26$, 38, and 46 days. 
Clearly, the bremsstrahlung emission from SN Ia models is far below the observed 
luminosity from SN~2005ke.

A more likely source of the X-ray emission is circumstellar interaction,
probably with mass lost by the companion.
The reported velocity of the Si\,II 6,355\AA\ absorption line
minimum ($13,300~{\rm km~s}^{-1}$; Patat \& Baade 2005) 
provides a lower limit to the shock velocity of the interaction.
Models of hydrodynamic interaction for a typical SN Ia density structure
show that the forward shock velocity may be $v_s \gs 40,000~{\rm km~s}^{-1}$
(Chevalier \& Fransson, in preparation).
Assuming a constant mass loss rate $\dot{M}$ and wind velocity $v_{\rm w}$ from
the progenitor's companion, the thermal X-ray luminosity of the forward shock region is
$L_{\rm x} = 1/(\pi m^2) \Lambda(T) \times (\dot{M}/v_{\rm w})^2\times(v_{\rm s} t)^{-1}$
(Immler, Wilson \& Terashima 2002)\footnote{Note the missing factor 4 as a 
correction to Immler, Wilson \& Terashima (2002)}, 
where $m$ is the mean mass per particle ($2.1\times10^{-24}$~g for a H+He plasma) 
and $\Lambda (T)$ the cooling function of the heated plasma at temperature $T$. 
Because of the higher density at the reverse shock, it is likely for the reverse
shock component to dominate the luminosity by a factor of $10$--$100$ (Chevalier 1982).
Adopting $L_{\rm reverse} = 30 \times L_{\rm forward}$, an effective cooling function 
of $\Lambda_{0.3-2} = 3 \times 10^{-23}~{\rm ergs~cm}^3~{\rm s}^{-1}$ for an optically
thin thermal plasma with temperature of $T = 10^9$~K for the forward shock (Fransson, 
Lundqvist \& Chevalier 1996; Raymond, Cox \& Smith 1976), and 
$v_s = 40,000~{\rm km~s}^{-1}$, a mass-loss rate of 
$\dot{M} \approx 3 \times 10^{-6}~M_{\odot}~{\rm yr}^{-1}~(v_{\rm w}/10~{\rm km~s}^{-1})$ 
with an uncertainty of a factor of 2--3 is inferred.
Assuming different plasma temperatures in the range $10^6$--$10^9$ K would lead to 
changes in the emission measure of $\ls 40\%$. 

The mass-loss rate is one of the lowest reported for any SN 
progenitor system detected either in X-rays (Immler \& Lewin 2003) or in the radio 
(Weiler et~al.\ 2002). However, Panagia et al.\ (2006) have recently presented 
upper limits on the radio emission from a number of SNe Ia which are interpreted 
as setting upper limits on $\dot{M}$ as low as 
$\sim3\times 10^{-8}~M_{\odot}~{\rm yr}^{-1}~(v_{\rm w}/10~{\rm km~s}^{-1})$.
In addition, Soderberg (2006) reports the nondetection of SN~2005ke at 8.46 GHz,
although flux limits are not given. 

X-ray emission from SNe is usually interpreted as thermal radiation.
At low circumstellar densities, nonthermal mechanisms may dominate
and $\dot M$ could be as low as $10^{-7}~M_{\odot}~{\rm yr}^{-1}$ for 
$v_{\rm w}=10~{\rm km~s}^{-1}$ in SN 2005ke (Chevalier \& Fransson in preparation). 
However, the nondetection of radio emission could limit the applicability of 
these mechanisms.

We rebinned 268~ks of XRT data of SN~2005ke into five consecutive time bins with 
similar exposure times (47, 47, 62, 56, and 55~ks) to further study the temporal
behavior of the X-ray emission (see Fig.~3, left-hand panel). Although the X-ray 
light curve 
is consistent with a constant X-ray luminosity during the period monitored, we 
find marginal evidence that the X-ray luminosity might have increased during the
early observations. The X-ray rise could be due to decreasing absorption by 
material along the line of sight to the hot gas as the expanding shell becomes 
optically thin, although we do not expect local absorption to be important for
our deduced parameters.

Significant excess in the UV output of SN~2005ke starts around day 35 after the 
explosion, especially in the $UVW1$ and $UVW2$ filters (see Fig.~3, right-hand panel). 
No such excess is observed 
in the optical band lightcurves. Possible effects of CSM interaction on UV emission
are inverse Compton scattering of photospheric photons by hot electrons or a 
reduction of UV line blanketing due to heating and ionization by circumstellar 
radiation. The luminosity of the UV excess dropped by a factor of $>4$  
between days 45 and 105, as estimated from comparisons 
with the other SN Ia light curves. If the X-ray luminosity experienced   
a similar drop during that time period, the \C\ upper limit may not be in 
conflict with the XRT detection. 



Evidence for CSM interaction has only been reported for a few SNe Ia 
(e.g., SNe 1999ee, Mazzali et~al.\ 2005; 2002ic, Hamuy et~al.\ 2003, Chugai, Chevalier
\& Lundqvist 2004, Kotak et~al.\ 2004, Deng et~al.\ 2004, Nomoto et~al.\ 2005; 
2003du, Gerardy et~al.\ 2004; 2005cg, Quimby et~al.\ 2006; and 2005gj, 
Prieto et~al.\ 2005) based on optical spectroscopy. Our detection of CSM interaction 
in X-rays is independently confirmed by the excess of UV emission after maximum. 
The most plausible explanation for the presence of dense 
CSM is that a more massive companion star ($\ls 10~M_{\odot}$) blew-off its outer 
layers via stellar winds or Roche lobe overflow over the last few ten thousand 
years before the accreted mass from the companion star pushed the progenitor close 
to the Chandrasekhar limit, resulting in a SN Ia explosion. 

\acknowledgments            
We gratefully acknowledge support provided                                     
by DOE's Scientific Discovery through 
Advanced Computing Program grant DE-FC02-01ER41189 (L.S.T.),
STSci grant HST-GO-10182.75-A (P.A.M.), 
NASA  Chandra Postdoctoral Fellowship grant PF4-50035 (D.P.),
and NSF grant AST-0307366 (R.A.C.).
We wish to thank H. Tananbaum for approving a \C\ DDT request.


\clearpage

\begin{deluxetable}{lcccccccccc}
\tabletypesize{\scriptsize}
\tablecaption{{\sl Swift} X-Ray Telescope Observations of Type Ia Supernovae \label{tab1}}
\tablewidth{0pt}
\tablehead{
\colhead{SN} &
\colhead{Galaxy} &
\colhead{Start} &
\colhead{Stop} &
\colhead{Distance} &
\colhead{$N_{\rm H}$} &
\colhead{Exposure} &
\colhead{Rate} &
\colhead{$f_{\rm x}$} &
\colhead{$L_{\rm x}$} &
\colhead{$\dot{M}$} \\
& 
& 
\colhead{[MJD]} &
\colhead{[MJD]} &
\colhead{[Mpc]} &
\colhead{[$10^{20}$]} &
\colhead{[ks]} &
\colhead{[$10^{-4}$]} &
\colhead{[$10^{-14}$]} &
\colhead{[$10^{39}$]} &
\colhead{[$10^{-5}$]} \\
\noalign{\smallskip}
\colhead{(1)}  &
\colhead{(2)}  &
\colhead{(3)}  &
\colhead{(4)}  &
\colhead{(5)}  &
\colhead{(6)}  &
\colhead{(7)}  &
\colhead{(8)}  &
\colhead{(9)}  &
\colhead{(10)} &
\colhead{(11)}
}
\startdata
2005am & NGC 2811	& 53433.94  & 53507.14	& \phantom{0}33.6	& 4.5 
	& \phantom{0}62.6  
	& $<6.2$ & \phantom{0}$<3.6$	& \phantom{0}\phantom{0}$<4.9$ 
	& \phantom{0}$<1.7$ \\
2005cf & MCG -01-39-3	& 53525.04  & 53574.67	& \phantom{0}27.4	& 7.0
	& \phantom{0}53.8  
	& $<3.5$ & \phantom{0}$<2.2$	& \phantom{0}\phantom{0}$<1.9$ 
	& \phantom{0}$<1.0$ \\
2005df & NGC 1559	& 53592.11  & 53711.25	& \phantom{0}18.4	& 2.7 
	& \phantom{0}25.8  
	& $<9.8$ & \phantom{0}$<5.4$	& \phantom{0}\phantom{0}$<2.2$ 
	& \phantom{0}$<1.4$ \\
2005ke & NGC 1371	& 53688.68 & 53799.72	& \phantom{0}20.7	& 1.4 
	& 267.8	
	& $2.2\pm0.6$ & $0.4\pm0.1$ 
	& \phantom{0}\phantom{0}\phantom{0}$0.2\pm0.1$ 
	& $0.3\pm0.1$ \\
2005gj & anonymous	& 53698.13  & 53698.33	& 264.5	& 7.1 & 
	\phantom{0}\phantom{0}5.0   
	& $<35.0$ & $<21.3$	& $<1,790$
	& $<37$ \\
2005hk & UGC 272	& 53677.75  & 53744.66	& \phantom{0}55.6	& 2.8
	& \phantom{0}45.1
	& $<4.1$ & \phantom{0}$<2.2$	& \phantom{0}\phantom{0}$<8.3$
	& \phantom{0}$<1.7$ \\
2006E  & NGC 5338	& 53748.83  & 53763.23	& \phantom{0}11.5	& 2.1 
	& \phantom{0}10.1 
	& $<13.8$ & \phantom{0}$<7.4$	& \phantom{0}\phantom{0}$<1.2$
	& \phantom{0}$<0.8$ \\
2006X  & NGC 4321	& 53774.34  & 53806.77	& \phantom{0}17.1	& 2.4 
	& \phantom{0}25.3  
	& $<7.8$ & \phantom{0}$<4.3$	& \phantom{0}\phantom{0}$<1.5$
	& \phantom{0}$<0.7$
\enddata
\tablecomments{
(1)~Name of the SN;
(2)~host galaxy;
(3)~start of the observation in Modified Julian Day;
(4)~end of the observation in Modified Julian Day;
(5)~distance in units of Mpc, using the NED redshifts of the host galaxies
    and assuming $H_0 = 71~{\rm km~s}^{-1}~{\rm Mpc}~^{-1}$,
    $\Omega_{\Lambda}=2/3$,  $\Omega_{\rm M}=1/3$; 
(6)~Galactic foreground column density in units of $10^{20}~{\rm cm}^{-2}$,
    Dickey \& Lockman 1990;
(7)~\S\ XRT exposure time in units of ks;
(8)~$3\sigma$ upper limit to the 0.2--10~keV count rate in units of 
    $10^{-4}~{\rm cts~s}^{-1}$;
(9)~$3\sigma$ upper limit to the 0.2--10~keV X-ray band flux in units of 
    $10^{-14}~{\rm ergs~cm}^{-2}~{\rm s}^{-1}$;
(10)~$3\sigma$ upper limit to the 0.2--10~keV X-ray band luminosity in units of 
    $10^{39}~{\rm ergs~s}^{-1}$;
(11)~mass-loss rate of the progenitor systems in units of 
    $10^{-5}~M_{\odot}~{\rm yr}^{-1}$.
}
\end{deluxetable}


\clearpage

\begin{figure}
\plotone{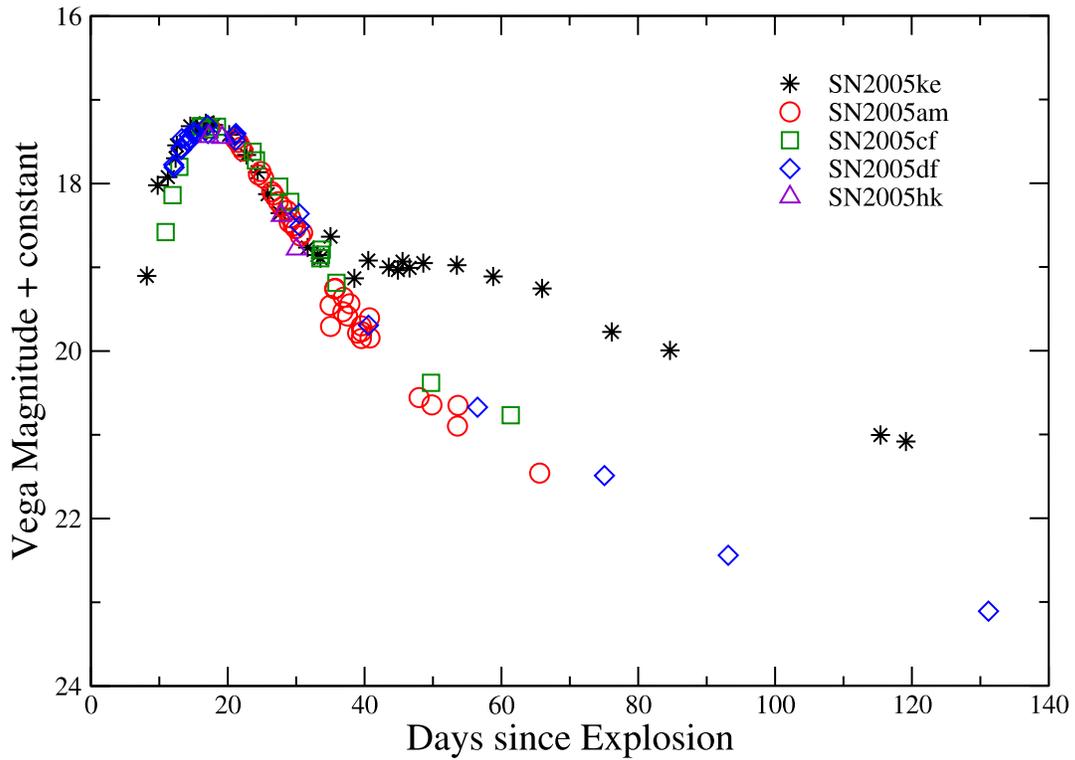}
\caption{Comparison of the $UVW1$ magnitudes of SNe Ia observed with \S. 
The lightcurves of SNe 2005am, 2005cf, 2005df, and 2005hk are shifted 
vertically to align them with the SN~2005ke template near maximum. 
The time is given in days after the outburst. 
The excess in the UV emission of SN~2005ke starting around day 35 after 
the outburst is likely caused by CSM interaction.
\label{f1}}
\end{figure}

\clearpage

\begin{figure}
\plotone{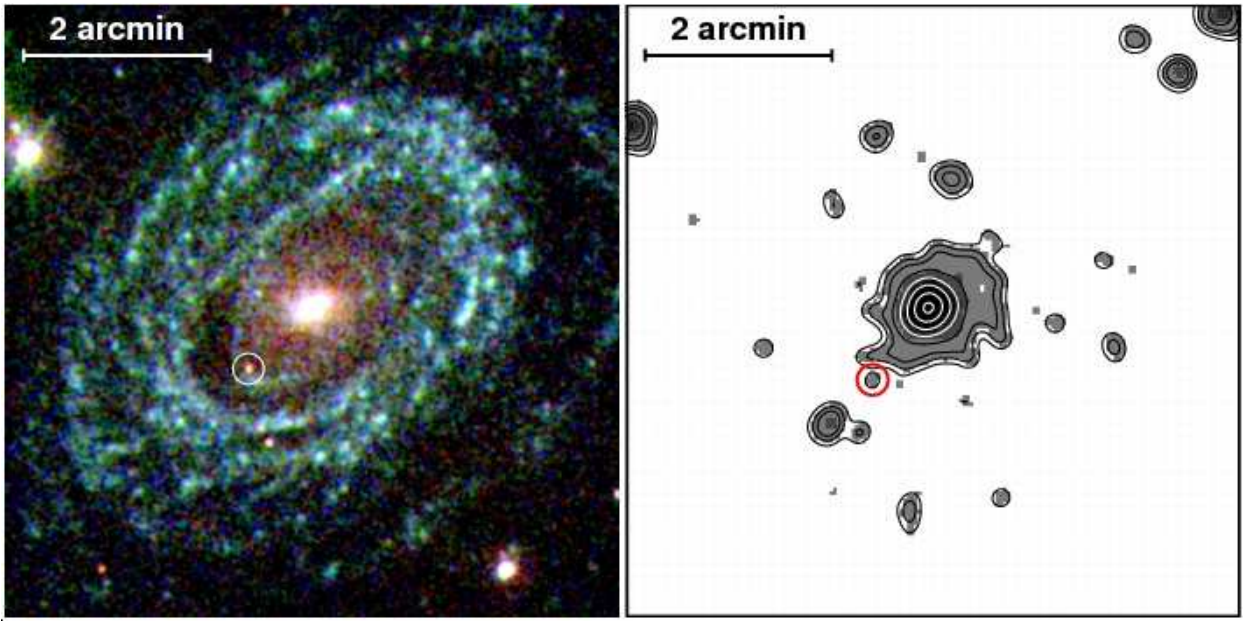}
\caption{\S\ UV and X-ray image of SN~2005ke and its host galaxy NGC 1371. 
{\bf Left-hand panel:} The UV image was constructed from the UVOT $UVW1$ 
(750 s exposure time; red), $UVW2$ (1,966 s; green) and $UVM2$ (1,123 s; blue) 
filters obtained on 2005-11-14.67~UT and slightly smoothed with a Gaussian filter 
of 1.5 pixel (FWHM). The position of SN~2005ke is indicated by a white circle of 
$10''$ radius. 
{\bf Right-hand panel:} The (0.2--10 keV) X-ray image was constructed from 
the merged 268 ks XRT data and is slightly smoothed with a Gaussian filter of 
1.5 pixels (FWHM). The position of SN~2005ke is indicated by a red circle of $10''$ 
radius. Contour levels are 0.3, 0.6, 0.9, 1.5, 3, 6, 12, and 20 counts 
${\rm pixel}^{-1}$. Same scale as the UV image.
\label{f2}}
\end{figure}

\clearpage

\begin{figure}
\plottwo{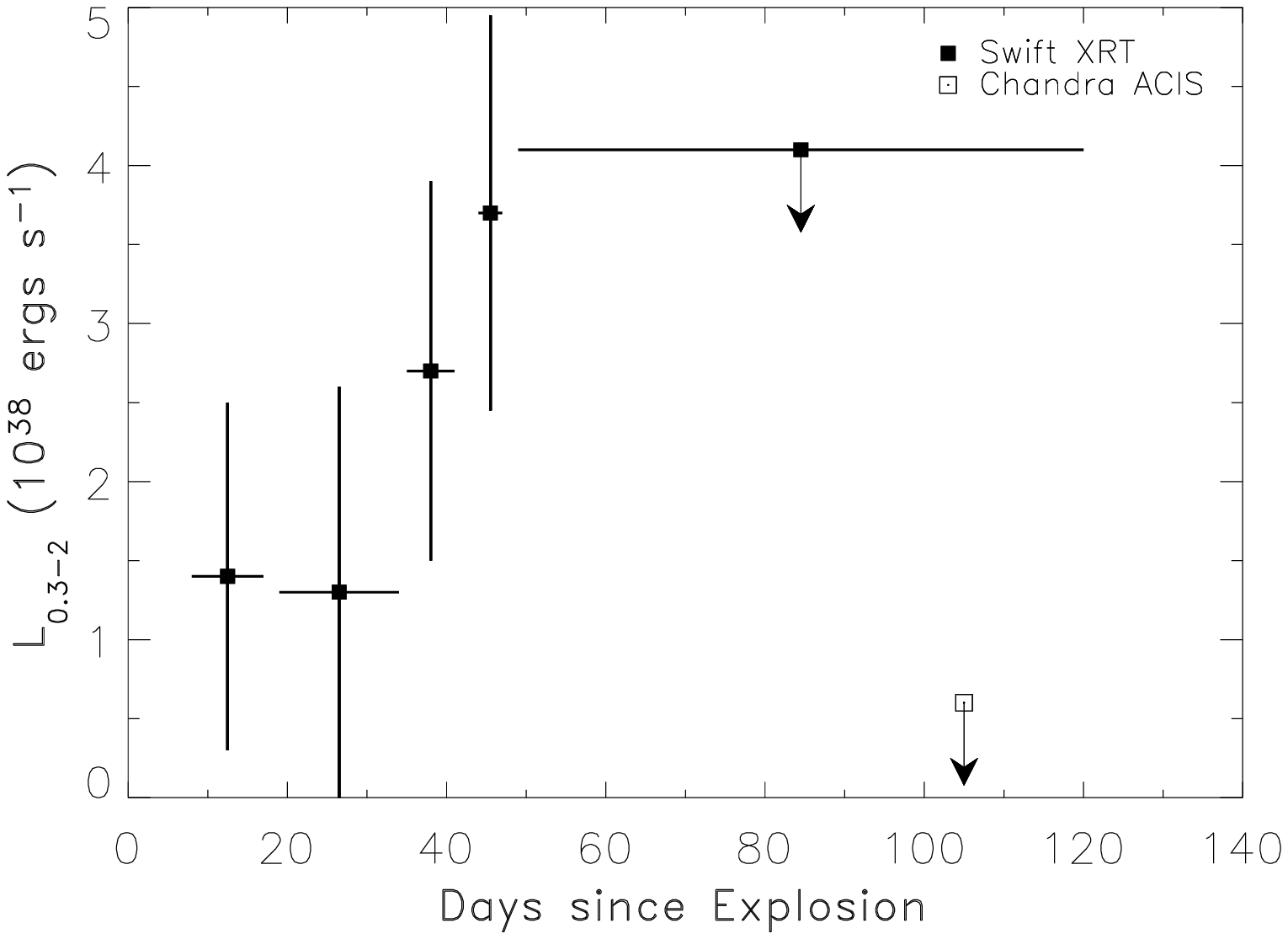}{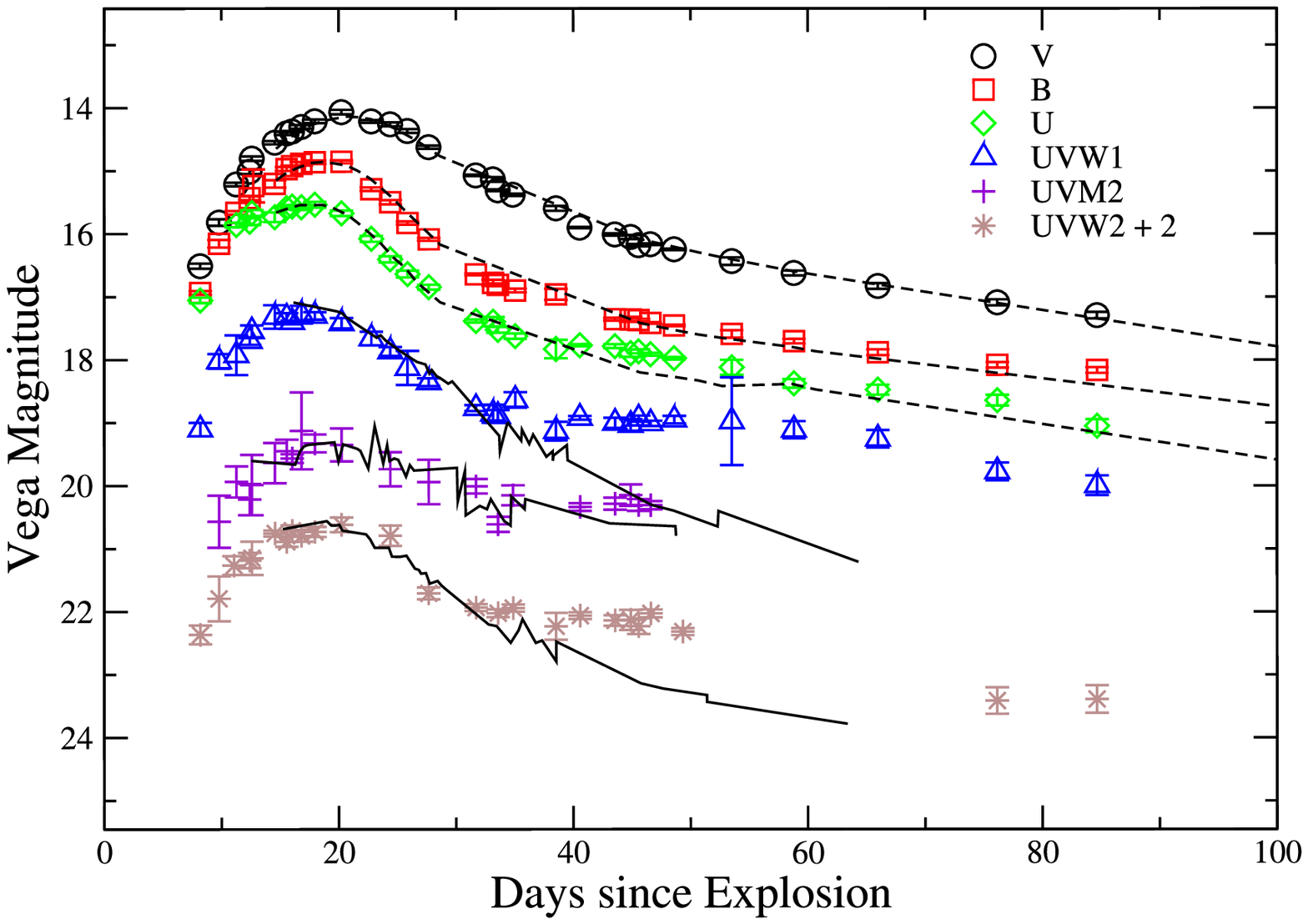}
\caption{{\bf Left-hand panel:} 
Soft (0.3--2~keV) X-ray band lightcurve of SN~2005ke as observed with the 
\S\ XRT. The time is given in days after the outburst (2005-11-06). 
The average X-ray luminosity over the observed period of 8--120 days after 
the explosion is $L_{0.3-2} = (2.0\pm0.5) \times 10^{38}~{\rm ergs~s}^{-1}$. 
Vertical error bars are statistical $1\sigma$ errors; horizontal error bars
indicate the periods covered by the observations (which are not contiguous).
The upper limit is at a $3\sigma$ level of confidence.
{\bf Right-hand panel:} Optical/UV lightcurves of SN~2005ke in all six 
UVOT filters. For comparison, the $V$, $B$, and $U$ lightcurves of the 
Type Ia SN~1999by (Garnavich et~al.\ 2004) and the UV ($UVW1$, $UVM2$, $UVW2$) 
lightcurves of SN~2005am (Brown et~al.\ 2005) are drawn as lines shifted vertically 
to match the SN~2005ke lightcurves near maximum.
\label{f3}}
\end{figure}

\end{document}